\documentclass[12pt,letterpaper]{article}  

\usepackage[includeheadfoot, 
            marginratio={1:1,2:3}, 
            width=412pt, 
            height=688pt,]{geometry}

\usepackage{amsmath}
\usepackage{amsfonts}
\usepackage{amssymb}

\usepackage{graphicx}
\usepackage{subfigure}
\usepackage{empheq}
\usepackage[margin=10pt,font=small]{caption}

\newcommand{\nc}{\newcommand}
\nc{\beq}{\begin{equation}}
\nc{\eeq}{\end{equation}}
\nc{\bea}{\begin{eqnarray}}
\nc{\eea}{\end{eqnarray}}
\def\ov{\overline}
\def\IR{\mathbb{R}}

\newcommand{\eq}[1]{\begin{equation}
                     \begin{split} #1 \end{split}
                     \end{equation}}

\newcommand{\tri}{\hspace{-3.5pt}\vartriangle\hspace{-3.5pt}}
\newcommand{\quat}{\hspace{-1.2pt}\diamond\hspace{-1.2pt}}

\begin{document}

\vspace*{-1.5cm}
\begin{flushright}
  {\small
  MPP-2010-128\\
  ITP-UU-10/36\\
  SPIN-10/31\\ 
  NSF-KITP-10-127
  }
\end{flushright}

\vspace{1.5cm}
\begin{center}
  {\LARGE
Nonassociative Gravity in String Theory? \\[0.2cm]
  }
\end{center}

\vspace{0.75cm}
\begin{center}
  Ralph Blumenhagen$^{1,3}$ and Erik Plauschinn$^{2,3}$
\end{center}

\vspace{0.1cm}
\begin{center} 
\emph{$^{1         }$ Max-Planck-Institut f\"ur Physik (Werner-Heisenberg-Institut), \\ 
   F\"ohringer Ring 6,  80805 M\"unchen, Germany } \\[0.1cm] 
\vspace{0.1cm} 
\emph{$^{2}$ Institute for Theoretical Physics and Spinoza Institute, \\
Utrecht University, 3508 TD  Utrecht, The Netherlands}  \\[0.1cm] 
\vspace{0.1cm} 
\emph{$^{3}$ Kavli Institute for Theoretical Physics, Kohn Hall, \\
  UCSB, Santa Barbara, CA 93106, USA} \\

\vspace{0.2cm}

 \vspace{0.5cm} 
\end{center} 

\vspace{1cm}


\begin{abstract}
In an on-shell conformal field theory approach,
we find indications of a three-bracket structure for target 
space coordinates in general closed string backgrounds.
This generalizes the appearance of noncommutative gauge theories
for open strings in two-form backgrounds to a  putative
non\-com\-mu\-ta\-tive/non\-asso\-ciative gravity theory
for closed strings probing curved backgrounds with non-vanishing
three-form flux.
Several aspects and consequences of the three-bracket structure are
discussed and a new type of generalized  uncertainty principle  is
proposed.
\end{abstract}

\clearpage


\section{Introduction}
\label{sec:intro}

String theory provides a unified framework for both quantum 
gravity and quantum gauge  theories, that is for all
interactions we observe in nature, where gravity arises from
closed strings while gauge theories are due to open strings.
It is well known that gauge theories can become ultraviolet
finite via the procedure of renormalization, however
in order to provide an ultraviolet finite theory of gravity
one expects the notion of a smooth space-time to 
break down  at very short distances.
Noncommutative (NC) geometry provides
a mathematical framework to describe such  space-times.

In view of this general expectation, it appears somewhat surprising
that noncommutative geometry first appeared in a clear way 
on the boundary of  open strings attached to a D-brane
carrying  non-trivial two-form 
flux \cite{Connes:1997cr,Chu:1998qz,Schomerus:1999ug}. 
This string theoretical result corresponds to the quantum mechanical
Landau quantization of cyclotron orbits of  a charged
particle in a constant magnetic field.  
Therefore, it is  the
gauge theory which becomes noncommutative  \cite{Seiberg:1999vs} and not
the gravity theory. 
The origin of this noncommutativity for the end-points of
open strings can be traced back to the
fact that on the 
boundary of a disk, one can define an 
ordering of two points close to each other.
Inserting  vertex operators and introducing a non-trivial background which 
is sensitive to the ordering -- such as a constant two-form flux --
 can lead to noncommutativity.
On the other hand, the closed string analogue is clearly different as
here two vertex operators are inserted in the bulk of 
a two-sphere $S^2$ and no unambiguous ordering can be defined.
Therefore, one does not expect the same kind of noncommutativity to arise.
(See \cite{Frohlich:1993es} for a proposal of closed string NC geometry, and
more recently  \cite{Lust:2010iy}.)
These statements can be checked explicitly by computing the equal-time, 
equal-position commutator 
$\lim_{\sigma'\to \sigma} [X^\mu (\sigma,\tau), X^\nu(\sigma',\tau)]$
for open and for closed strings.

However, if  one considers  {\em three} nearby points  on the world-sheet
$S^2$ of a closed string, one can 
very well decide whether the loop connecting the three points has positive
or negative orientation. Thus, if there exists a background field
which distinguishes these two orientations, one would expect a non-vanishing
result not for the simple commutator, but  for
the cyclic double commutator
\begin{equation}
\label{jacobiid}
\begin{split}
  \bigl[X^\mu,X^\nu,X^\rho\bigr]:=
  \lim_{\sigma_i\to \sigma}\; \bigl[ [X^\mu (\sigma_1,\tau),
    X^\nu(\sigma_2,\tau)],X^\rho(\sigma_3,\tau)\bigr] +{\rm cyclic}\;. 
\end{split}
\end{equation}
To get a better understanding of this expression, let us mention that for a fundamental product $x^i \bullet x^j$ one 
can define a three-bracket as
\begin{equation}
\begin{split}
   \bigl[x^1,x^2,x^3\bigr] = \sum_{\sigma\in P_3} {\rm sign}(\sigma) \Bigl(
   \bigl(x^{\sigma(1)}\bullet x^{\sigma(2)}\bigr)\bullet x^{\sigma(3)} 
  - x^{\sigma(1)}\bullet \bigl(x^{\sigma(2)}\bullet x^{\sigma(3)}\bigr) \Bigr)\;,
\end{split}
\end{equation}\pagebreak[2]
being the  completely anti-symmetrized 
associator of this $\bullet\,$-product. 
Note that for an associative product, this expression vanishes 
which is also known as the Jacobi-identity. Therefore, a non-vanishing
cyclic double commutator \eqref{jacobiid} indicates that a $\bullet\,$-product 
is both noncommutative and nonassociative (NCA).

\bigskip
In this article, we pursue  a direct attempt to establish a non-trivial three-bracket 
for a certain class of treatable examples.\footnote{Such a three-bracket
has been discussed in M-theory for open membranes  moving in a constant
$C_3$-form background \cite{Hoppe:1996xp,Ho:2007vk,Chu:2009iv}. Moreover, algebras with Lie-type 
three-brackets
have  appeared in the formulation of an effective field theory
on a stack of M2-branes \cite{Bagger:2006sk}.} 
A natural candidate for a background field
leading to such a structure in the closed string sector is the three-from field strength
$H$.  
The main technical challenge is 
that via Einstein's equation, a non-trivial $H$-flux 
induces a curvature of space, so that
in general we have no means to solve the resulting non-linear
sigma model (NLSM) explicitly. However,  WZW-models \cite{Witten:1983ar}  
are known 
to describe compactifications on
group manifolds with non-trivial  $H$-flux \cite{Gepner:1986wi}. 
For the simplest case of $SU(2)$, the model is equivalent to a closed
string propagating on $S^3$ with $H$-flux. 
This is the prime example of a bosonic string compactification
that we will elaborate on  in this paper.
A second challenge is that when employing the framework of 
conformal field theory,
computations are done on-shell. Therefore, the background 
is guaranteed to satisfy
the string equations of motions and certain structures
might not be  visible (or indistinguishable from known results) since they vanish (or agree with familiar results) 
on-shell.

In the following discussion it will become clear that
our arguments are based on one
technical assumption, leading to (at least) three possible interpretations
of our result. 
We discuss each of these possibilities in some detail, however, to reach a conclusive picture further investigation is necessary.

\bigskip
This paper  is organized as follows.
In section \ref{sec_open}, we recall some features of open 
strings ending on a D-brane  endowed with background two-form flux.
We will be brief and restrict ourselves to aspects 
important in the course of this paper.
In section \ref{sec:closed}, we discuss the $SU(2)_k$ WZW model and its geometric
non-linear sigma model interpretation. 
In section \ref{sec:clna}, we carry out in detail a conformal
field theoretic computation of
a cyclic double commutator, which is to be considered 
the central part of this article.
In section \ref{sec:interpret}, we discuss some of the formal and conceptual 
consequences of our result from a wider perspective, and
present a proposal for a {\it generalized}
nonassociative gravity theory based on a three-bracket for the left- as well as for the 
right-moving sector of the closed string. 
Section \ref{sec:concl} contains our conclusions. Some important technical details are collected in the appendix, where in particular
 a quantum mechanical derivation of a new type
of uncertainty relation resulting from a non-vanishing 
three-bracket can be found.

\vfill


\section{Open String Noncommutativity}
\label{sec_open}

To illustrate the analogy between the open string and our upcoming discussion for the closed string, 
let us first review  the quantization of an open string 
in a  background with constant two-form flux. Here, we follow the discussion
in \cite{Chu:1998qz} where more details can be found.

We consider an open string with both endpoints on a D$p$-brane
carrying  constant two-form flux ${\cal F}_{ij}=B_{ij}+F_{ij}$, where $i,j=0,\ldots,p$.
This leads to mixed Neumann-Dirichlet boundary conditions 
longitudinal to the brane, so that the mode expansions for the 
corresponding free bosons read
\begin{equation}
\label{openexpand}
X^i(\sigma,\tau)=x^i_0+ \bigl(\alpha^i_0 \,\tau - \alpha^j_0 {\cal F}_j{}^i
    \sigma\bigr) + \sum_{n\ne 0} \frac{e^{-in\tau}}{n}\Bigl(
        i\,\alpha^i_n \cos (n\sigma) - \alpha^j_n {\cal F}_j{}^i \sin (n\sigma)\Bigr)\,.
\end{equation}
Here we normalized $0\le \sigma\le \pi$, and indices of $\mathcal F_{ij}$ are raised by the inverse metric $\eta^{ij}={\rm diag}\,(-1,+1,\ldots,+1)$. As carried out in \cite{Chu:1998qz}, 
the commutation relations for  the modes appearing in \eqref{openexpand}
can be obtained via canonical quantization.
Using these relations,  the equal-time
commutator is evaluated as
\begin{equation}
\label{commut}
  \bigl[X^i(\sigma_1,\tau),X^j(\sigma_2,\tau)\bigr] =-2i\hspace{1pt}\alpha'  \big(M^{-1}{\cal F}\bigr)^{ij}
  \left[ P(\sigma_1,\sigma_2)  + \sum_{n\ne 0} \frac{\sin
      n(\sigma_1+\sigma_2)}{n}\right] ,
\end{equation}
where $M_{ij}=\delta_{ij}-{\cal F}_i{}^k {\cal F}_{kj}$ and matrix products are understood.
The function $P$ is a continuous 
linear expression in the world-sheet coordinates  $\sigma_i$ of the form
\eq{
\label{open_p}
P(\sigma_1,\sigma_2)= \sigma_1+\sigma_2 -\pi \;,
}
which arises purely from the commutation relations involving the zero modes $x_0^i$ and $\alpha_0^i$. The sum in \eqref{commut} originates  from the oscillator modes $\alpha_n^i$ for $n\neq0$, and can be further evaluated using the Fourier transform
\begin{equation}
\label{gammala}
\gamma(\varphi) =\sum_{n=1}^\infty  \:\frac{\sin (n\varphi)}{ n}=
\left\{ 
\begin{array}{@{\hspace{2pt}}c@{\hspace{15pt}}l}
\frac{1}{2} (\pi-\varphi) &  0<\varphi<2\pi\;, \\[1.5mm]
0&  \varphi=0,2\pi \; . 
\end{array}
\right.
\end{equation}
Coming back to the commutator, using equations \eqref{open_p} and \eqref{gammala}, we see that for  $0< \sigma_1+\sigma_2 < 2\pi$ 
the two terms in \eqref{commut} cancel.
However, 
on the boundaries $\sigma_1=\sigma_2=0$ and $\sigma_1=\sigma_2=\pi$
one obtains
\begin{equation}
   \label{comm_open_res}
   \bigl[X^i(0,\tau),X^j(0,\tau)\bigr]=-\bigl[X^i(\pi,\tau),X^j(\pi,\tau)\bigr]=
    2\pi i\hspace{1pt}\alpha'  \bigl(M^{-1}{\cal F}\bigr)^{ij}\; .
\end{equation}
In summary, the equal-time, equal-position commutator between two target-space
coordinates $X^i(\sigma,\tau)$ does not vanish along a  D-brane carrying  non-trivial
two-form flux $\mathcal F_{ij}$. 
For later purpose, let us emphasize two points of this computation:
\begin{itemize}

\item Even without
knowing  the zero mode contribution  $P(\sigma_1,\sigma_2)$ explicitly, 
we could have guessed this function by requiring 
the commutator \eqref{commut} to vanish for generic points on the world-sheet.
In turn, the non-zero result in \eqref{comm_open_res} arises from the boundaries of the open string  due to the discontinuity of $\gamma(\varphi)$ at $\varphi=0\!\mod2\pi$.

\item Since the equal-time, equal-position commutator \eqref{comm_open_res} is
independent of the world-sheet coordinates $\sigma$ and $\tau$, one can indeed
conclude that this world-sheet  computation reveals a feature of the
target space (as probed by an open string).

\end{itemize}

\begin{figure}[t]
\centering	
\includegraphics[width=0.485\textwidth]{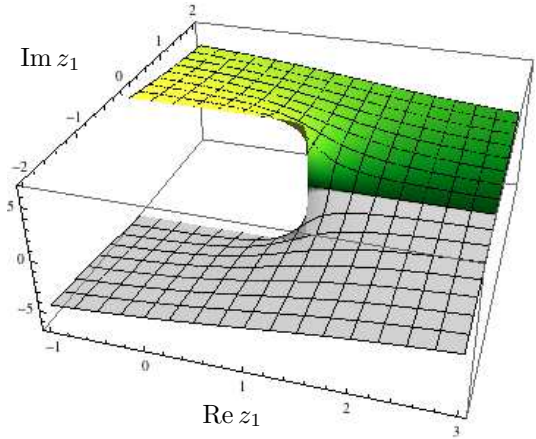}
\begin{picture}(0,0)
\put(-130,7){\footnotesize ${\rm Re}\,z_1$}
\put(-199,142){\footnotesize ${\rm Im}\,z_1$}
\end{picture}
\caption{The imaginary part of the two-point function \eqref{two_point} for $z_2=1$ and $\alpha'=2$, $f=1$. Note that the coordinates $z_i$ are defined only on the upper half-plane, but the lower half-plane ${\rm Im}\,z_1<0$ is included for illustrative purposes.}
\label{fig1}
\end{figure}

Another way to detect the noncommutative nature of the setting above is to
consider the two-point function of two fields $X^i(\sigma,\tau)$. In
particular, in the presence of  a constant two-form flux $\mathcal F_{12}=f$ we compute for instance \cite{Callan:1986bc,Abouelsaood:1986gd}
\begin{equation}
  \label{two_point}
   \bigl\langle X^1 (z_1)\, X^2(z_2) \bigr\rangle =  
   \alpha' \: \frac{f}{1+f^2}\: \log\left( \frac{z_1-\overline z_2}{\overline z_1 - z_2} \right)\; ,
\end{equation}
where in order to work on  an Euclidean world-sheet we have 
performed a Wick rotation $\tau_i\to i\,\tau_i$ 
and  introduced $z_i = \exp(\tau_i + i \sigma_i)$. As illustrated in figure \ref{fig1}, the function \eqref{two_point}  has a jump when changing the order of $z_1$ and $z_2$ on the real line, which indicates the noncommutativity. 

To conclude this section, we note that 
the result of a two-form flux inducing  noncommutativity of brane coordinates
is completely general, and has also been studied for co-dimension
one branes in the $SU(2)$ WZW model \cite{Alekseev:1999bs}. 
However, due to a background $H$-flux in this case,
 it turns out that the obtained structure is not only noncommutative but also
nonassociative \cite{Alekseev:1999bs,Cornalba:2001sm,Herbst:2001ai}.


\section{Closed Strings on Curved Spaces}
\label{sec:closed}

Motivated by the results for the open string, our strategy now is to compute the expression
\eqref{jacobiid} for the closed string. However, since we want to consider non-trivial $H$-flux, 
we have to work
on curved background spaces on which we can introduce only
local coordinates.


\subsubsection*{The $\mathbf{SU(2)}$ WZW Model}

Let us start our discussion by considering  the WZW model for the group manifold
$SU(2)$. The corresponding action is given by
\begin{equation}
\label{wzwnaction}
\begin{split}
    S=\hphantom{+}&\:\frac{k}{16\pi} \int_{\partial \Sigma} d^2x \, {\rm Tr}\, \Bigl[ (\partial_\alpha g)(\partial^\alpha g^{-1}) \Bigr]
       \\ 
 -&\:\frac{ik}{24\pi} \int_\Sigma d^3y \, \epsilon^{\tilde\alpha\tilde\beta\tilde\gamma} \:
     {\rm Tr}\,\Bigl[ (g^{-1} \partial_{\tilde\alpha} g)(g^{-1} \partial_{\tilde\beta} g)(g^{-1} \partial_{\tilde\gamma} g) \Bigr]\;,
\end{split}
\end{equation}
where $k\in\mathbb Z^+$ denotes the level and 
 $\Sigma$ is a three-dimensional manifold with boundary $\partial\Sigma$. 
The indices take values $\alpha=1,2$ and $\tilde\alpha,\ldots =1,2,3$, which are raised or lowered by the metrics $h_{\alpha\beta}={\rm diag}(+1,+1)$ and $h_{\tilde \alpha\tilde \beta}={\rm diag}(+1,+1,+1)$, respectively.
Parametrizing an element $g\in SU(2)$ as
\begin{equation}
  \label{def_g}
     g=\left(\begin{matrix} e^{i\eta^2} \cos\eta^1 &  e^{i\eta^3} \sin\eta^1 \\
          -e^{-i\eta^3} \sin\eta^1 &  e^{-i\eta^2}
          \cos\eta^1 \end{matrix}\right) \;,
\end{equation}
with $0\le \eta^1\le \pi/2$, $0\le \eta^{2,3}\le 2\pi$, the first term in
\eqref{wzwnaction} can be written as 
\eq{
     S_{\rm kin.}=\frac{k}{8\pi} \int_{\partial\Sigma} d^2x 
   \Bigl[ \:\partial_\alpha \eta^1\, \partial^{\alpha} \eta^1 + (\cos\eta^1)^2\,
     \partial_\alpha \eta^2\, \partial^{\alpha} \eta^2 
+(\sin\eta^1)^2\, \partial_\alpha \eta^3\, \partial^{\alpha} \eta^3 \:\Bigr] .
\label{wzwnaction_kin}
}
Note that \eqref{wzwnaction_kin} 
is  a non-linear sigma model (in conventions $\alpha'=2$) 
with target space $S^3$, where the latter is given in Hopf coordinates $\eta^i$ with metric
\begin{equation}
\label{metricsp}
ds^2=k\, \Bigl[ (d\eta^1)^2 +  (\cos\eta^1)^2\, (d\eta^2)^2
+(\sin\eta^1)^2 \,(d\eta^3)^2\Bigr]\; .
\end{equation}
By comparing with the usual metric on $S^3$, one  infers that the radius of the three-sphere is given by $R=\sqrt k$.
The second  term in the WZW model \eqref{wzwnaction} can be expressed as 
\begin{equation}
    S_{\rm WZ}= -\frac{ik}{12\pi} \int_{\Sigma} d^3y\, \epsilon_{ijk}\, \epsilon^{\tilde\alpha\tilde\beta\tilde\gamma} \sin\eta^1 \,
    \cos\eta^1 \; \partial_{\tilde\alpha} \eta^i\,    \partial_{\tilde\beta} \eta^j\, \partial_{\tilde\gamma} \eta^k \;,
\end{equation}
which corresponds to a background flux 
\eq{
  \label{h-flux_01}
  H=-2k \sin\eta^1 \cos\eta^1 \, d\eta^1\wedge d\eta^2\wedge d\eta^3 \;.
}
To summarize, the $SU(2)$ WZW model at level $k$ (with central charge
$c=3k/[k+2]$) is equivalent to the non-linear
sigma model on  $S^3$ with radius $R=\sqrt{k}$ 
and background flux \eqref{h-flux_01}.

Recall also that  in a string theory context, this configuration, together with a 
linear
dilaton background of  central charge $c=1+\frac{6}{k}+\mathcal O(k^{-2})$, 
describes the deep throat limit
of an NS five-brane geometry \cite{Callan:1991at}.  
Furthermore, it is remarkable that the metric $G$
and the $B$-field are not changed due to higher order $\alpha'$-corrections.
Indeed, as shown in \cite{Khuri:1990mg}, 
utilizing that $S^3$ is   parallelizable,
such corrections only re-adjust
the dilaton while leaving $G$ and $B$ at their tree level values.
However, in this paper, we mostly focus on the
WZW part and will only comment on the dilaton in section \ref{sec:interpret}.


\subsubsection*{Conserved Currents and Kac-Moody Algebras}

Solving the model \eqref{wzwnaction} directly in terms of Hopf coordinates $\eta^i$
is not easily possible, but  it is well known that the
WZW model  actually is exactly solvable. To see this, we 
introduce a complex coordinate $z=\exp(x^1+ix^2)$ and
define the currents 
\begin{equation}
\label{def_j}
J  = 
J^a \, \frac{\sigma^a}{\sqrt 2}= - k\, \bigl( \partial_z g \bigr)\,   g^{-1}\, ,
\hspace{40pt}
\ov J  = 
\ov J{\vphantom J}^a \, \frac{\sigma^a}{\sqrt 2}= + k\, g^{-1} \bigl( \partial_{\ov z} g \bigr) \;.
\end{equation}
Note that here and in the following, $\sigma^a$ with $a=1,2,3$ are the Pauli matrices and  summation over repeated indices is understood.
From the equation of motion of the WZW model \eqref{wzwnaction} it follows that the currents $J^a$ are holomorphic and that the $\ov J{\vphantom J}^a$ are anti-holomorphic.
Therefore, one  can perform the Laurent expansions 
\begin{equation}
\label{expansion}
    J^a(z)=\sum_{n\in \mathbb Z} j^a_n\, z^{-n-1} \;, \hspace{60pt}
    \ov J{\vphantom J}^a(\ov z)=\sum_{n\in \mathbb Z} \ov j{\vphantom j}^a_n\, \ov z^{-n-1} \;.
\end{equation}
The symmetry transformations of the WZW model then translate into the following commutation relations  for the modes $j^a_n$ and $\ov j{\vphantom j}^a_n$ 
\begin{equation}
  \label{km_alg}
  \begin{split}
   \bigl[j^a_m,j^b_n\bigr]\hspace{0.75pt}&=  i f^{ab}{}_c\, j^c_{m+n} + k\, m\, \delta_{m+n}\,
   \delta^{ab}\; , \\[2mm]
    \bigl[\ov j{\vphantom j}^a_m,\ov j{\vphantom j}^b_n\bigr]&=  i f^{ab}{}_c\, \ov j{\vphantom j}^c_{m+n} 
    + k\, m\, \delta_{m+n}\,  \delta^{ab} \;,
\end{split}
\hspace{60pt}\bigl[j^a_m,\ov j{\vphantom j}^b_n\bigr]=0 \;,
\end{equation}
which define two independent Kac-Moody algebras. Note that the structure constants for $SU(2)$ in our convention read $f^{abc}=\sqrt{2} \,\epsilon^{abc}$, and indices are raised or lowered by $\delta^{ab}$ and $\delta_{ab}$, respectively.

To study the properties of the WZW model in more detail, let
us employ the parametrization \eqref{def_g} in \eqref{def_j} and express 
 the two  currents \eqref{expansion}  as follows
\begin{equation}
\label{strombein}
  J^a(z)= - i\sqrt{2}\,k\,  {\rm E}^a{}_i\,  \partial_z \eta^i, \hspace{60pt}
  \ov J{\vphantom J}^a(\ov z)= - i\sqrt{2}\,k\, \ov {\rm E}{\vphantom {\rm E}}^a{}_i\,  \partial_{\ov z} \eta^i\;,
\end{equation}
where a summation over $i=1,2,3$ is understood. The matrices ${\rm E}$ and $\ov{\rm E}$ depend on
$\eta^1$ as well as on $\eta_\pm^{23}=\eta^2\pm \eta^3$ and
are given by
\begin{equation}
\label{thematrix}
 {\rm  E} = 
 \left( \begin{matrix} \sin\eta_+^{23} & -\sin\eta^1\cos\eta^1\cos\eta_+^{23}& \sin\eta^1\cos\eta^1\cos\eta_+^{23} \\
  \cos\eta_+^{23} & \sin\eta^1\cos\eta^1\sin\eta_+^{23}& -\sin\eta^1\cos\eta^1\sin\eta_+^{23} \\
  0 & (\cos\eta^1)^2 & (\sin\eta^1)^2
  \end{matrix} \right) \;,
\end{equation}
and
\begin{equation}
\label{thematrixb}
  \ov{\rm  E} =
 \left( \begin{matrix} \sin\eta_-^{23} & -\sin\eta^1\cos\eta^1\cos\eta_-^{23}& -\sin\eta^1\cos\eta^1\cos\eta_-^{23} \\
  -\cos\eta^{23}_- & -\sin\eta^1\cos\eta^1\sin\eta_-^{23}& -\sin\eta^1\cos\eta^1\sin\eta_-^{23} \\
  0 & -(\cos\eta^1)^2 & (\sin\eta^1)^2
  \end{matrix} \right)\; .
 \end{equation}


\subsubsection*{Geometric Interpretation}

We now turn to a geometric interpretation of the above setting. In particular, the matrices $\rm E$ and $\ov {\rm E}$ can be used to define two three-beins
\begin{equation}
\label{threebeins}
  e^a=  \sqrt{k} \, {\rm E}^a{}_i\,  d \eta^i, \hspace{60pt}
  \ov e^a=  \sqrt{k}\, \ov {\rm E}\vphantom{\rm E}^a{}_i\,  d \eta^i\; ,
\end{equation}
which diagonalize the metric \eqref{metricsp}, that is
$ds^2=\sum_{a} e^a\otimes e^a=\sum_{a} \ov e^a\otimes \ov e^a$. 
The corresponding vector fields are given by
\begin{equation}
\label{vectorfields}
  e_a=  \frac{1}{\sqrt{k}} \: {\rm E}_a{}^i\,  \partial_{\eta^i}, \hspace{60pt}
  \ov e_a=  \frac{1}{\sqrt{k}}\: 
  \ov {\rm E}\vphantom{\rm E}_a{}^i
  \,  \partial_{\eta^i}\; ,
\end{equation}
where ${\rm E}_a{}^i$ denotes the inverse transpose of \eqref{thematrix} and similarly for $\ov {\rm E}\vphantom{\rm E}_a{}^i$. These vector fields satisfy commutation relations
\begin{equation}
\label{commrelac}
  \bigl[e_a,e_b\bigr]=C_{ab}{}^c\,  e_c \;, \hspace{60pt}
  \bigl[\ov e_a,\ov e_b\bigr]=\ov C_{ab}{}^c\,  \ov e_c \;, 
\end{equation}
with structure constants $C_{ab}{}^c=\ov C_{ab}{}^c=\sqrt{2/k}\,f_{ab}{}^c$.
Returning to the three-beins \eqref{threebeins}, by explicit computation we find that 
\begin{equation}
\label{cartananw}
  de^a+ \frac {1}{\sqrt{2k}}\, f^a_{\ bc}\,  e^b\wedge e^c=0 \;, \hspace{60pt}
  d\ov e^a+  \frac{1}{\sqrt{2k}}\, f^a_{\ bc} \,  \ov e^b\wedge \ov e^c=0\; .
\end{equation}
With the help of  Cartan's structure equations $de^a+\omega^a{}_b\wedge e^b = T^a$ and $d\omega^a{}_b+\omega^a{}_c\wedge \omega^c{}_b = R^a{}_b$, one identifies and computes for the {\em torsion-free} connection that
\begin{align}
  \label{connection_free_01}
  \omega^a{}_b \hspace{8.5pt}&= - \frac{1}{\sqrt{2k}}\: f^a{}_{bc} e^c \;,\hspace{30pt} 
  &\ov \omega^a{}_b \hspace{8.5pt}&=  - \frac{1}{\sqrt{2k}}\: f^a{}_{bc} \ov e^c \;, \\
  T^a \hspace{12pt}&= 0 \;,  &\ov T\vphantom{T}^a \hspace{12pt}&= 0 \;, \\
  \label{curvature}
  R^a{}_{bcd} &=  +\frac{1}{2k} \: f^a{}_{bp} f^p{}_{cd} \;, 
  &\ov R\vphantom{R}^a{}_{bcd} &=  +\frac{1}{2k} \: f^a{}_{bp} f^p{}_{cd} \;.
\end{align}
In addition, the $H$-flux \eqref{h-flux_01} can be expressed in terms of
the three-beins \eqref{threebeins} as
\begin{equation}
  \label{h-flux_02}
   H=  -\frac{2}{ \sqrt{k}}\, e^1\wedge e^2\wedge e^3 
   =+\frac{2}{ \sqrt{k}}\, \ov e^1\wedge \ov e^2\wedge \ov e^3   \;.
\end{equation}
Let us emphasize the important technical detail of the
same signs in \eqref{connection_free_01} for $e^a$ and $\ov e^a$  reflecting
a  left-right symmetric coupling of the metric in the string
action. 
On the other hand, the opposite signs in \eqref{h-flux_02} show that the
$B$-fields couples in a left-right asymmetric fashion.
(As usual, left refers to the holomorphic and right to the anti-holomorphic part.)

Coming back to the exact solvability of the WZW model, geometrically this is
related to the fact that one can define  {\em torsion-full} connections
in terms of \eqref{connection_free_01} and $H_{abc}$  as
\begin{equation}
    \Omega^{+\,a}{}_b =\omega^a{}_b +  \frac{1}{2} H^a{}_{bc} e^c\;, \hspace{60pt}
    \ov\Omega\vphantom{\Omega}^{-\,a}{}_b = \ov\omega^a{}_b -  \frac{1}{2} H^a{}_{bc} \ov e^c\;,     
\end{equation}
where $H_{abc}$ can be deduced from $H=\frac{1}{3!} H_{abc} e^a\wedge e^b\wedge e^c$. For these connections one finds
\begin{align}
  \label{torsion_458}
  \mathcal T^+{}_{abc}  \hspace{4.5pt}&= - H_{abc} \;, &
  \ov{\mathcal{T}}{\vphantom{\mathcal T}}^-{}_{abc} \hspace{4.5pt}& = + H_{abc} \;, \\
  {\cal R}^+{}_{abcd}&= \frac{2}{ k}\left( f_{abu} f^u{}_{cd} + f_{cau} f^u{}_{bd} +
                                    f_{bcu} f^u{}_{ad}\right) =0 \;,\hspace{30pt}
  &    \ov{\cal R}\vphantom{\cal R}^-{}_{abcd}&\equiv0\; ,    
  \label{curvbb}                                
\end{align}
where for the vanishing of $\mathcal R^+$ the Jacobi identity was employed, and $\ov{ \mathcal R}\vphantom{\cal R}^-$ vanishes identically.
Note that since $S^3$ is parallelizable, it was expected that there indeed exist  connections with vanishing curvature.


\section{Closed String Nonassociativity}
\label{sec:clna}


In the previous section, we have mainly reviewed the well-known geometry for
the exactly solvable $SU(2)_k$ WZW model.
However,  let us now introduce fields   $X^a(z,\ov z)$ according to 
\begin{equation}
\label{free_coords_X_01}
\begin{split}
  J^a(z)&= - i\,\sqrt{k}\,\partial_z X^a(z,\ov z)= - i\sqrt{2} \,k\,  {\rm
    E}^a{}_i(\vec \eta)\,
  \partial_z \eta^i(z,\ov z)  \;,\\
  \ov J\vphantom{J}^a(z)&= -i\,\sqrt{k}\,\partial_{\ov z} X^a(z,\ov z)= 
  -i\sqrt{2} \,k\,  {\rm \ov E}\vphantom{\rm E}^a{}_i(\vec \eta)\,
  \partial_{\ov z} \eta^i(z,\ov z) \;,
\end{split}
\end{equation}
where we have indicated that the matrices  ${\rm E}^a{}_i$ and
${\rm \ov E}\vphantom{\rm E}^a{}_i$ depend on the target space
coordinates $\eta^i$. 
It is clear that
the $X^a$ do not correspond to  bona fide global coordinates
on $S^3$ since there does not exist a flat metric on $S^3$.
However, as shown in \cite{Braaten:1985is},  if the $X^a(z,\ov z)$ satisfy their  
(free) equations of motion, the $\eta^i(z,\ov z)$ do  so as well.

\pagebreak[2]

In the present section we  compute
the cyclic double
commutator  of three {\em local} coordinates $X^a$ at a specific point on the target space.
Since the double commutator  is a local quantity,  we can imagine to probe 
the geometry around a  point $\vec \eta_0$ on a  three-sphere $S^3$
by a closed string. 
Writing then
\eq{
  \label{free_coords_X_02}
  X^a(z,\ov z)=X^a(z)+\overline{X}\vphantom{X}^a(\overline z) 
}
and using \eqref{free_coords_X_01},
locally we can identify the
left- and right-moving coordinates  as
\begin{equation}
\label{free_coords_X_03}
\begin{split}
  X^a(z)\simeq  \sqrt{2k}\,   {\rm E}^a{}_i(\vec \eta_0)\, \eta^i(z)\;,
  \hspace{60pt}
  \ov X\vphantom{X}^a(\ov z)\simeq  \sqrt{2k}\,  {\rm \ov E}\vphantom{\rm E}^a{}_i (\vec
  \eta_0)\, \ov \eta^i(\ov z)\; .
\end{split}
\end{equation}
The mode expansions of $X^a(z)$ and $\ov X\vphantom{X}^a(\ov z)$ are found 
by  integrating  the expansions of the currents given in
 \eqref{expansion}. In particular, for the holomorphic part we arrive at 
\begin{equation}
\label{xexpand}
   X^a(z)=\frac{i}{\sqrt k}\: x_0^a -  \frac{i}{ \sqrt{k}} \,j^a_0 \log z
    + \frac{i}{  \sqrt{k}}\, \sum_{n\ne 0}
    \frac{j^a_n}{n}\: z^{-n} \;,
\end{equation}
and a similar expression is obtained for the anti-holomorphic part $\ov
X\vphantom{X}^a(\ov z)$. The modes $j_n^a$ in \eqref{xexpand} satisfy the
corresponding Kac-Moody algebra given in \eqref{km_alg}, however, a priori it
is not clear what the precise form of the commutation relations involving
$x_0^a$ is. In the following, we are going to fix this contribution in analogy
to the open string as in section \ref{sec_open}.


\subsubsection*{Cyclic Double Commutator}

Let us consider the cyclic double commutator for the holomorphic part $X^a(z)$ of the  free fields \eqref{free_coords_X_02}
\eq{
  \label{double_123}
  \bigl[X^a(z_1),X^b(z_2),X^c(z_3) \bigr]=
  \bigl[\hspace{1pt} \bigl[ X^a(z_1),X^b(z_2)\bigr]\,,\,X^c(z_3) \bigr] +{\rm cyclic}\;,
}
evaluated at equal times. For our choice of complex coordinates $z_i=\exp(\tau_i + i\sigma_i)$ this implies $|z_1|=|z_2|=|z_3|$, which will always be understood for the expression \eqref{double_123}.
To simplify the following formulae, let us furthermore introduce $\mathbf x^a$, $\mathbf p^a$ and $\mathbf j^a$ as 
\eq{
  \mathbf x^a = \frac{i}{\sqrt k}\: x_0^a \;, \hspace{34pt}
  \mathbf p^a(z) = -  \frac{i}{ \sqrt{k}} \,j^a_0 \log z \;,\hspace{34pt}
  \mathbf j^a(z) =  \frac{i}{  \sqrt{k}}\, \sum_{n\ne 0}
    \frac{j^a_n}{n}\: z^{-n} \;.
}
For the computation of \eqref{double_123}, we first collect all terms involving $\mathbf x^a$ into a so far undetermined function  $\mathcal P^{abc}$
\eq{
  \label{zm_contr_01}
  \mathcal P^{abc}(z_1,z_2,z_3) = 
  \bigl[\hspace{1pt}\mathbf x^a, \mathbf x^b, \mathbf x^c\hspace{1pt} \bigr]
  + \bigl[\hspace{1pt}\mathbf x^a, \mathbf x^b, \, \cdot\, \hspace{1pt} \bigr]
  + \bigl[\hspace{1pt}\mathbf x^a, \, \cdot\, , \,  \cdot\,  \hspace{1pt} \bigr] 
  + \ldots \;.
}
For all other contributions in \eqref{double_123}, we employ the Kac-Moody algebra \eqref{km_alg} of the modes $j^a_n$. In particular, we compute 
\eq{
  \bigl[\hspace{1pt}\mathbf p^a(z_1), \mathbf p^b(z_2), \mathbf p^c(z_3)\hspace{1pt} \bigr]
  \sim \bigl( f^{ab}{}_u f^{uc}{}_v + f^{bc}{}_u f^{ua}{}_v + f^{ca}{}_u f^{ub}{}_v \bigr) \, j_0^v = 0 \;,
}
which vanishes due to the Jacobi identity for the structure constants $f^{ab}{}_c$. In a similar fashion,  we obtain vanishing expressions for
\eq{
  \arraycolsep2pt
  \begin{array}{@{}cccccl@{}}
  \displaystyle \bigl[\hspace{1pt}\mathbf p^a(z_1), \mathbf p^b(z_2), \mathbf j^c(z_3)\hspace{1pt} \bigr]
  &+&
  \displaystyle  \bigl[\hspace{1pt}\mathbf p^a(z_1), \mathbf j^b(z_2), \mathbf p^c(z_3)\hspace{1pt} \bigr]
  &+&
  \displaystyle \bigl[\hspace{1pt}\mathbf j^a(z_1), \mathbf p^b(z_2), \mathbf p^c(z_3)\hspace{1pt} \bigr]
  &= 0 \;, 
  \\[2mm]
  \displaystyle \bigl[\hspace{1pt}\mathbf p^a(z_1), \mathbf j^b(z_2), \mathbf j^c(z_3)\hspace{1pt} \bigr]
  &+&
  \displaystyle \bigl[\hspace{1pt}\mathbf j^a(z_1), \mathbf p^b(z_2), \mathbf j^c(z_3)\hspace{1pt} \bigr]
  &+&
  \displaystyle \bigl[\hspace{1pt}\mathbf j^a(z_1), \mathbf j^b(z_2), \mathbf p^c(z_3)\hspace{1pt} \bigr]
  &= 0 \;,
  \end{array}
}
where for the second line we utilized in addition to the Jacobi identity  that $f^{abc}=f^{bca}=f^{cab}$.
However, for the double commutator involving three $\mathbf j^a$, we find the non-trivial result 
\eq{
  \label{double_125}
  \bigl[\hspace{1pt}\mathbf j^a(z_1), \mathbf j^b(z_2), \mathbf j^c(z_3)\hspace{1pt} \bigr]
  = - \frac{f^{abc}}{\sqrt k} \sum_{\substack{n,m\neq 0 \\ n+m\neq 0}} \frac{1}{n\,m} \left( \frac{z_3}{z_1} \right)^n 
  \left( \frac{z_3}{z_2} \right)^m + {\rm cyclic}\;.
}
Remember that  this expression is understood to be evaluated at equal times. For the right-hand side in  \eqref{double_125}, we split the sums in the following way and compute
\eq{
  \label{double_126}
  \Gamma(\sigma_1,\sigma_2,\sigma_3)= & - \sum_{n,m\neq 0} \frac{1}{n\,m} \left( \frac{z_3}{z_1} 
 \right)^n  \left( \frac{z_3}{z_2} \right)^m
 - \sum_{n\neq 0} \frac{1}{n^2} \left( \frac{z_2}{z_1} 
 \right)^n  
  + {\rm cyclic} \\[1.8mm]
 =&  \left\{
  \begin{array}{@{}c@{\hspace{30pt}}l}
   -\pi^2 & \sigma_1=\sigma_2=\sigma_3 \;, \\[1.8mm]
    0
    & {\rm else} \;.
    \end{array}
    \right.
}
Combining  the above results, we arrive at an expression for the equal-time
double commutator
of the holomorphic fields $X^a(z)$ of the form
\eq{
  \label{holdoubleyeah}
  \bigl[X^a(z_1),X^b(z_2),X^c(z_3) \bigr]= \mathcal P^{abc}(z_1,z_2,z_3) 
  + \frac{f^{abc}}{\sqrt k}\:\Gamma(\sigma_1,\sigma_2,\sigma_3) \;.
}
For the computation in the anti-holomorphic sector, we note that the modes $\ov j\vphantom{j}^a_n$ satisfy the same Kac-Moody algebra as $j^a_n$. Furthermore, we have $\ov z_i = e^{\tau_i - i \sigma_i}$ and so we only need to replace $\sigma_i\to -\sigma_i$ in the result \eqref{holdoubleyeah} for the holomorphic sector. However, observe that the function $\Gamma$ is invariant under that substitution.
Therefore, the result for the full equal-time double commutator reads
\eq{
\label{doubleyeah}
  &\bigl[X^a(z_1,\ov z_1),X^b(z_2,\ov z_2),X^c(z_3,\ov z_3) \bigr]  \\
  &\hspace{80pt}=\mathcal P^{abc}(z_1,z_2,z_3) 
  + \ov{\mathcal{ P}}\vphantom{\mathcal P}^{abc}(\ov z_1,\ov z_2,\ov z_3) 
  + 2\:\frac{f^{abc}}{\sqrt k}\:\Gamma(\sigma_1,\sigma_2,\sigma_3) \;. 
}

\pagebreak[4]

\subsubsection*{Zero Mode Contribution and Final Result}

It is now tempting to follow the same logic as for the open string
computation. That is, we fix the unknown contribution $\mathcal P + \ov{\mathcal{P}}$ of the zero modes $x_0^a$ and $\ov x\vphantom{x}^a_0$ by:
\begin{center}
\label{assum_1}
\begin{tabular}{@{}p{70pt}p{320pt}@{}}
{\it Assumption}\hspace{1.5pt}: & The zero mode contribution $\mathcal P + \ov{\mathcal{P}}$ 
is {\it continuous}; and for the three points $z_i$ not all equal, the equal-time double commutator has to vanish.
\end{tabular}
\end{center}
More concretely, this assumption means that 
\begin{equation}
\label{qqsecond}
  \mathcal P^{abc}(z_1,z_2,z_3) 
  + \ov{\mathcal{ P}}\vphantom{\mathcal P}^{abc}(\ov z_1,\ov z_2,\ov z_3) 
  = 0\;.
\end{equation}
Using then \eqref{double_126} and \eqref{qqsecond} in \eqref{doubleyeah}, we arrive at the following expression for  the equal-time, equal-position cyclic double commutator\,\footnote{For ease of notation, when the dependence of the fields $X^a(z,\ov z)$ on the world-sheet coordinates $(z,\ov z)$ is omitted, the cyclic double commutator is understood to be evaluated at equal-time {\em and} equal-position.}  
\begin{equation}
  \label{result}
   \bigl[X^a,X^b,X^c \bigr] 
   := \lim_{z_i\to z} \bigl[X^a(z_1,\ov z_1),X^b(z_2,\ov z_2),X^c(z_3,\ov z_3) \bigr] 
   =  - \:\frac{2\pi^2}{\sqrt k}\:  f^{abc}\; .
\end{equation}
Therefore, pursuing the same reasoning as for the open string,
we are led to the intriguing result that the fields
$X^a$  satisfy a non-vanishing three-bracket\,\footnote{We are grateful to 
Chong-Sun Chu for the following comment: 
for a particle in the presence of a magnetic field  $B$ in three  space dimensions
a result similar  to \eqref{result} can be obtained. In particular, in this situation it is known that
the covariant derivatives  only  satisfy the Jacobi-identity up
to a term ${\rm div} (B)$, that is the cyclic double commutator does not vanish for a magnetic monopole.}, 
where the right-hand side is constant and proportional to
the $SU(2)$ structure constants $f^{abc}$. 
Let us make a few further comments:
\begin{enumerate}

\item As for the open string, the mathematical reason behind the result
{\eqref{result}} is the discontinuity of  
$\Gamma(\sigma_1,\sigma_2,\sigma_3)$ in \eqref{double_126}.
 
\item The equal-time, equal-position  double commutator is independent
of the world-sheet coordinates. Thus, it is  expected to
reflect a property of the target space (as probed by a closed string).

\item Recalling that the radius of the three-sphere is  $R=\sqrt k$, we
  realize that in the large radius limit   $R\to \infty$ the NCA effect vanishes.

\item We  also computed the single commutator 
$\lim_{z_i\to z} [X^a(z_1,\ov z_1),X^b(z_2,\ov z_2) ]$ 
and found it to be dependent on the world-sheet coordinates. We therefore conclude
that the fundamental, well-defined target space structure is a 
three-bracket. Clearly, if a two-bracket with Lie-algebra structure would be well-defined  then the three-bracket vanishes.
\label{remark_4}

\item Since in this section we have not referred to a specific property of
the group $SU(2)$, our computation generalizes to any WZW model, that is
to string theory on any group manifold. This includes for instance
 the exactly solvable pp-wave backgrounds  discussed in \cite{Nappi:1993ie,Dolan:2002px}.

\end{enumerate}


\subsubsection*{Three-Point Function}

Finally, relating to our discussion in section \ref{sec_open}, we 
show that the correlation function of three fields $X^a$ features a jump. 
Integrating the three-point functions of three currents $J^a(z)$
and $\ov J\vphantom{J}^a(\overline  z)$ we obtain 
\begin{equation}
\label{three-point_01}
\begin{split}
  \bigl\langle X^a(z_1, \overline z_1) \; X^b(z_2,\overline z_2)\; & X^c(z_3,\ov z_3) \bigr\rangle \\
  &= \frac{\pi^2}{\sqrt k}\: \frac{f^{abc}}{9}   \left[
   L \Bigl({\textstyle{\frac{z_{12}}{z_{13}}}}\Bigr)
   +L \Bigl({\textstyle{\frac{z_{13}}{z_{23}}}}\Bigr)
   +L \Bigl({\textstyle{\frac{z_{32}}{z_{12}}}}\Bigr)
   +{\rm c.c.} \right],
\end{split}    
\end{equation}
where $z_{ij}=z_i-z_j$  and where  $L(x)$ denotes the Rogers dilogarithm defined
as 
\begin{equation}
  L(x)=\frac{6}{ \pi^2} \left({\rm Li}_2(x) + \frac{1}{ 2} \log (x) \log(1-x)\right)\; ,
\end{equation}
with
${\rm Li}_2(x)=\sum_{n= 1}^{\infty} \frac{x^n}{ n^2}$.
Let us then choose for instance $z_2\in\mathbb R$, $z_3=0$ and study  the behavior
of \eqref{three-point_01}.
\begin{figure}[t]
\centering	
\subfigcapskip7.5pt
\subfigure[$z_2=1$]{
\includegraphics[width=0.45\textwidth]{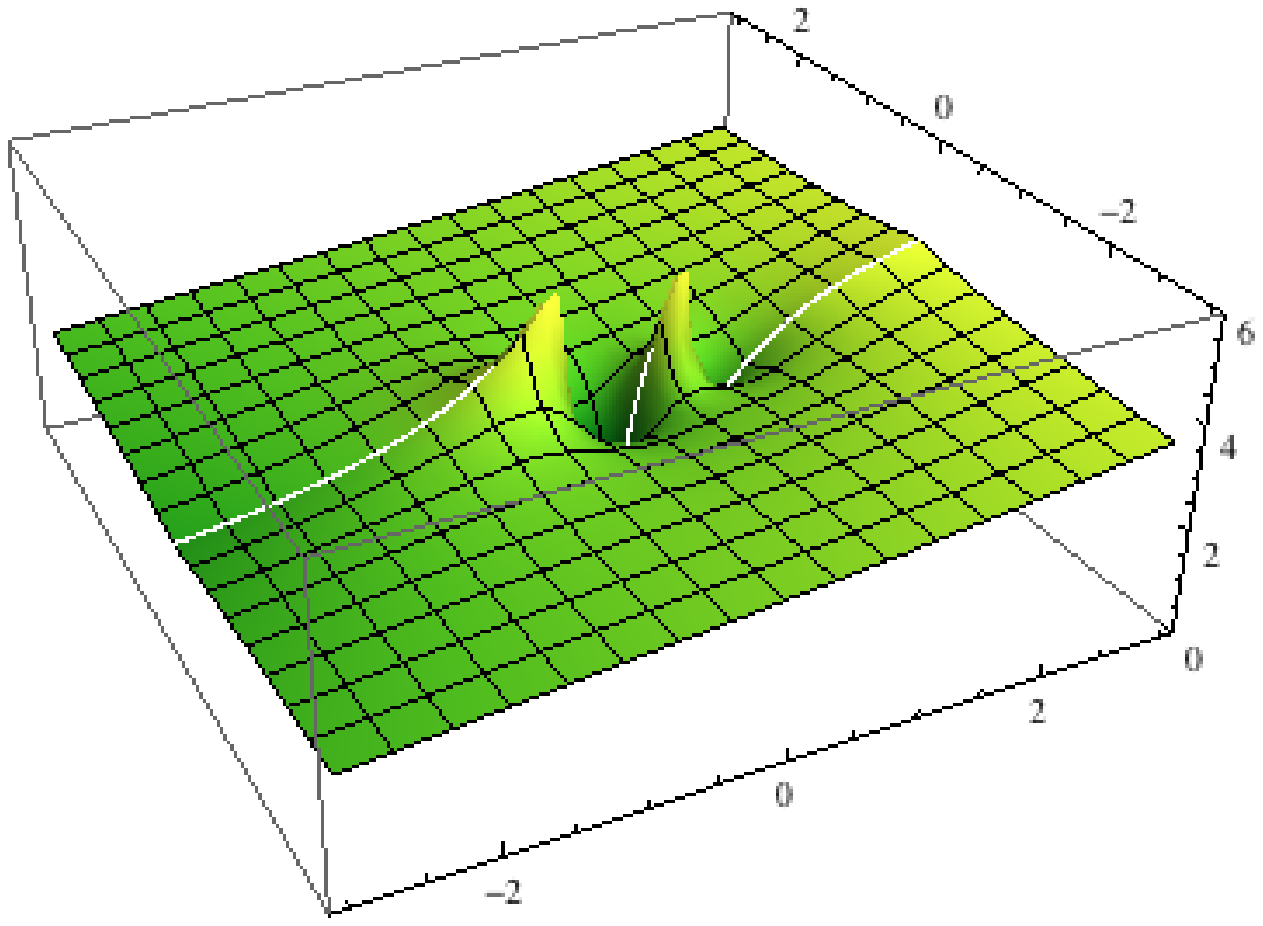}
\begin{picture}(0,0)
\put(-195,40){\footnotesize${\rm Im}\,z_1$}
\put(-82,5){\footnotesize${\rm Re}\,z_1$}
\end{picture}
\label{fig2_2}
}
\hfill
\subfigure[$z_2=10^{-4}$]{
\includegraphics[width=0.45\textwidth]{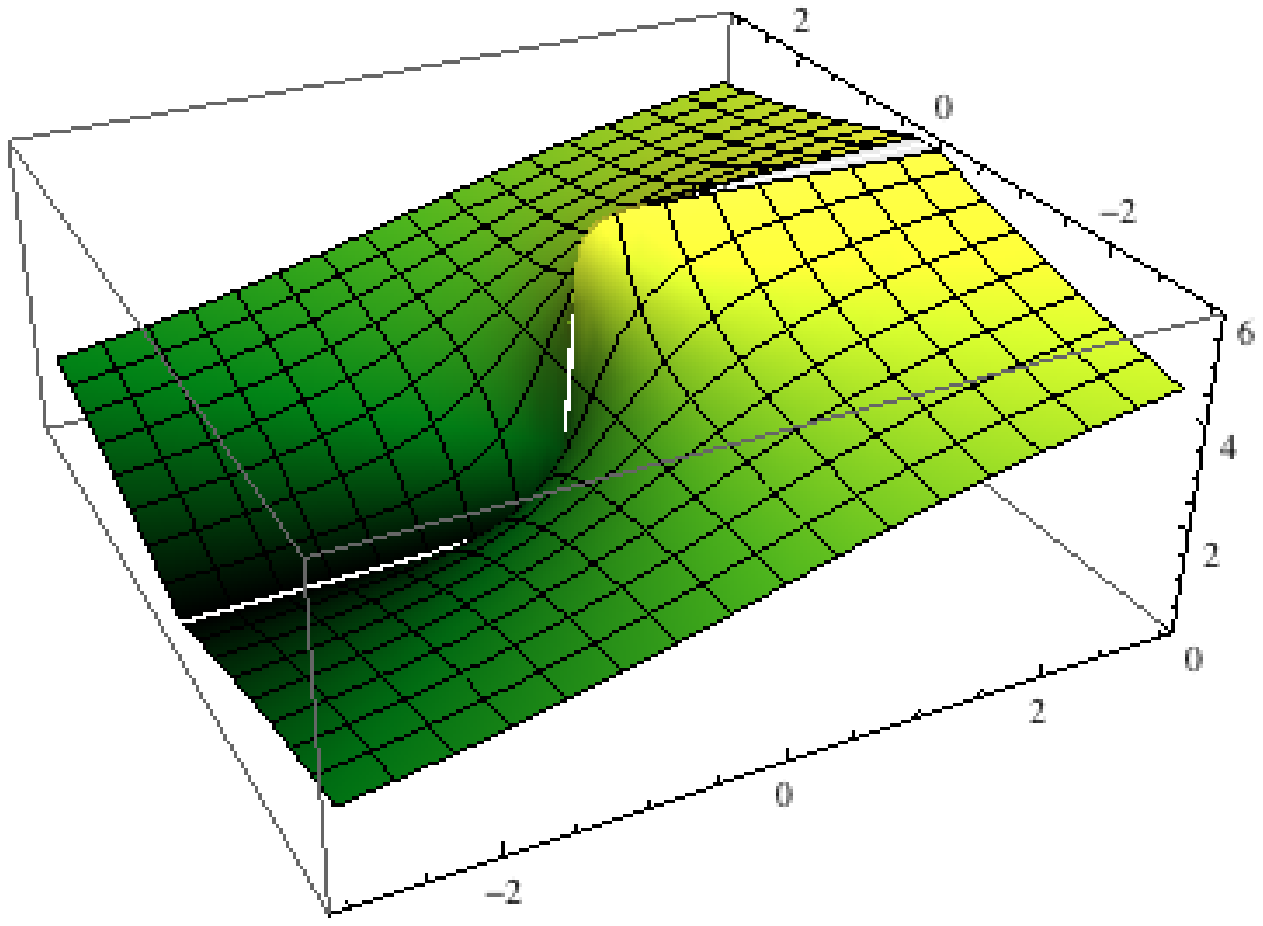}
\begin{picture}(0,0)
\put(-195,40){\footnotesize${\rm Im}\,z_1$}
\put(-82,5){\footnotesize${\rm Re}\,z_1$}
\end{picture}
\label{fig2_1}
}
\caption{\small{Behavior of three-point function \eqref{three-point_01} for $f^{123}=\sqrt{2}$, $k=1$, $z_3=0$ and different values of $z_2$. In the plots the dependence  on  $z_1$ is shown.}}
\label{fig2}
\end{figure}
As shown  in figure \ref{fig2_2}, for $z_2=1$ away from $z_3=0$ the three-point function features discontinuities when $z_1\to z_2=1$ and when $z_1\to z_3=0$. However, 
as illustrated in  figure \ref{fig2_1}, when all three points in \eqref{three-point_01} approach each other, that is when $z_1\to z_3=0$ and  $z_2\to z_3=0$, the three-point function develops a jump. 
We can therefore conclude that a non-vanishing double
commutator is related to jumps in the three-point
function when all three points coincide. This is analogous to the open string case where a jump in the two-point function indicates a non-vanishing commutator.


\section{Speculations about the Three-bracket}
\label{sec:interpret}

In this section, we present possible interpretations 
of the non-vanishing three-bracket \eqref{result}. 
This result was derived for the fields $X^a$, 
which were obtained by integrating the Kac-Moody currents \eqref{def_j}, and 
it is quite remarkable that even though the Laurent modes $j^a_n$ and $\ov j\vphantom{j}^a_n$ of the Kac-Moody currents satisfy a Jacobi-identity,  the equal-time, equal-position cyclic double
commutator of three   $X^a$ does not vanish.

More concretely, our aim is to identify  the  source of the
non\-com\-mu\-ta\-tive/non\-asso\-ciative structure obtained in the previous section.
Since we are working on-shell, this is  somewhat ambiguous. We therefore
discuss three possibilities and postpone a conclusive answer to 
future studies \cite{deser}.  
To start, we first  recall that the three-bracket \eqref{result}, indicating the NCA nature of our setting, is proportional to the $SU(2)$ structure constants and identify three quantities 
in the three-bein  basis  \eqref{threebeins}
which contain these $f^{abc}$:
\begin{enumerate}
\label{list}

\item The structure constants  $C^{abc}$ 
appearing in the Lie-algebra \eqref{commrelac}.

\item The torsion-free  spin connection given in equation \eqref{connection_free_01}.

\item The components  $H^{abc}$ of the $H$-field shown in  \eqref{h-flux_02}.

\end{enumerate}
For bona fide  coordinates on the three-sphere,  for instance the Hopf coordinates $\eta^i$ introduced in \eqref{def_g}, 
each of these possibilities would  lead to a different proposal  for the
origin of the NCA structure.
However, unfortunately we are not able to quantize the system, compute the cyclic
double commutator and detect a non-trivial three-bracket  in the geometric basis directly.


\subsubsection*{NCA Source: Structure Constants $C^{abc}$}

In the first case in the list above, the three-bracket in a geometric basis would be
proportional to $C^{ijk}$ which vanishes since the geometric vector fields
commute, that is $[\partial_{\eta^i},\partial_{\eta^j}]=0$.
Therefore, we would not expect any new structure in the geometric basis.


\subsubsection*{NCA Source: Connection $\omega^{abc}$}

In the second case, the three-bracket would be proportional to the  
completely anti-symmetrized part of the torsion-free connection
$\omega^{abc}$.
To make that more clear, we insert the local expressions
\eqref{free_coords_X_03} 
into the left- and right-moving three-brackets  for $X^a(z)$ and $\ov X{}^a(\ov z)$ and find
\begin{equation}
\label{threebrackgeo}
\begin{split}
  \bigl[\eta^i,\eta^j,\eta^k\bigr]_{\rm hol}  \simeq  -\frac{\pi^2}{2k^2} \,\frac{\epsilon^{ijk}}{\det ({\rm E})}\;,
  \hspace{50pt}
  \bigl [\ov\eta{}^i,\ov\eta{}^j,\ov\eta{}^k\bigr]_{\mbox{\scriptsize anti-hol}} 
  \simeq  -\frac{\pi^2}{2k^2} \,\frac{\epsilon^{ijk}}{\det (\ov{\rm E})}\;.
\end{split}
\end{equation}
Here we assume that $\eta^i(z,\ov z)=\eta^i(z)+\ov\eta{}^i(\ov z)$
and that three-brackets between mixed holomorphic $\eta^i(z)$ and anti-holomorphic fields $\ov\eta{}^i(\ov z)$ vanish. However, let us emphasize that a priori these assumptions are not clear.
Adding then both terms in \eqref{threebrackgeo} and noting that $\det ({\rm E})=-\det (\ov{\rm E})$, 
the full three-bracket for the geometric coordinates $\eta^i(z,\ov z)$ vanishes.
Since the corresponding Christoffel connection  is torsion-free and
thus its   totally antisymmetric part $\omega^{[ijk]}$ vanishes,  
this result would be consistent 
with the latter being the source for a non-vanishing three-bracket.
Note furthermore,  the sum $\eta^i(z,\ov z)=\eta^i(z)+\ov\eta{}^i(\ov z)$ is invariant under
the world-sheet parity transformation exchanging left- and right-moving coordinates.
Therefore, also the three-bracket for $\eta^i(z,\ov z)$ is expected to be invariant, that is the three-bracket should be a geometric  quantity associated with the metric.

Although this argument deserves further study, let us assume  it  to be correct and generalize the three-bracket \eqref{result} to any coordinate system (three-bein), not necessarily geometric, in the following way
\begin{equation}
\label{threealgg}
   \bigl[X^i,X^j,X^k\bigr]= \frac{\pi^2}{ \sqrt 2}\,   (\alpha')^2 
\, \omega^{ijk}\; ,
\end{equation}   
where $i,\ldots=1,2,3$. On dimensional grounds we have included the 
string scale  $\alpha'$ and have employed that $f^{abc} = -\sqrt{2k}\, \omega^{abc}$.
Note that generically the right-hand side of \eqref{threealgg} is coordinate
dependent and it remains to be seen how such a relation is to be understood
mathematically.

Analogous to the star-product for noncommutative geometry,
let us now introduce a three-product 
$F\tri G\tri H$ 
on the space of functions depending on the $X^i$ such that 
\eq{
\label{antisymtripcon}
   \bigl [X^i,X^j,X^k \bigr]=\sum_{\sigma\in P_3} {\rm sign}(\sigma) \;  
     X^{\sigma(i)}\tri X^{\sigma(j)}\tri X^{\sigma(k)} \; .
}
This anti-symmetrized expression is also called the Nambu-Heisenberg
commutator and one possibility to satisfy \eqref{antisymtripcon} is to choose the three-product as a generalization 
of the Moyal-Weyl star-product  in the following way
\eq{
\label{threebracketcon}
   F\tri G\tri H = \exp\left[ {\textstyle \frac{\pi^2}{6\sqrt{2}}} (\alpha')^2
     \, \omega^{ijk}\,
      \partial^x_{i}\,\partial^y_{j}\,\partial^z_{k} \right]\, F(x)\, G(y)\,
   H(z)\Bigr|_{x=y=z} \;.
}
The properties of such a product, in particular for  the case 
that $\omega^{ijk}$ is not completely antisymmetric or 
not constant, need to be understood better.
Let us also comment on the dilaton, in particular on the contribution of a non-trivial
dilaton gradient. We note that in \cite{Rey:1991uu} a more general connection $\Omega$ with torsion was considered, namely
\begin{equation}
      \Omega^{\pm\, abc}=\omega^{abc} + \delta^{c[a} \nabla^{b]}\Phi  \pm\frac{1}{ 2}
      H^{abc}\; ,
\end{equation}
where $\omega^{abc}$ are again the components of the spin connection.
These connections  $\Omega^\pm$ include  a non-trivial dilaton gradient,
so  it is conceivable that such a part has to be included
in the three-product \eqref{threebracketcon}.

\bigskip
Finally, we recall that open strings ending on a D-brane give rise to a gauge theory and that, in the presence of  background two-form flux on the D-brane, 
one can  define a  limit  (the so-called Seiberg/Witten limit)  in which this
gauge theory  is equivalent to a noncommutative gauge theory \cite{Seiberg:1999vs}. The product of
two functions for the latter is given by the 
star-product. 
Motivated by this observation for the open string, the question arises whether one can define an  
NCA version of Einstein-gravity for the closed string, such
that the higher order $\alpha'$-corrections  of the $\omega$ connection are
captured by the NCA three-product \eqref{threebracketcon}.\footnote{This might be similar to the  covariant gravity theory on the Moyal plane developed in \cite{Aschieri:2005yw}.} We will come back to this point on page \pageref{threebrackgeotwo}.


\subsubsection*{NCA Source: $H$-flux}

The third possibility  on page \pageref{list} for the origin of the NCA structure is 
the $H$-flux. This leads to the strongest proposal for the NCA source, as  
 this flux  does not  vanish  in any basis.

Relating to the discussion below equation 
\eqref{threebrackgeo}, let us observe that 
the antisymmetric combination of left- and right moving  fields 
$\tilde\eta^i(z,\ov z)=\eta^i(z)-\ov\eta^i(\ov z)$ yields the three-bracket
\begin{equation}
\label{hannover96}
\bigl[\tilde\eta^i,\tilde\eta^j,\tilde\eta^k\bigr]  \simeq  - \frac{\pi^2}{k^2} \,\frac{\epsilon^{ijk}}{\det ({\rm E})}\;,
\end{equation}
where we assumed again that three-brackets between mixed holomorphic $\eta^i(z)$ and anti-holomorphic fields $\ov\eta{}^i(\ov z)$ vanish.
Note that now the left-hand side in \eqref{hannover96} is odd under the world-sheet parity 
operation which exchanges left- and right-moving coordinates $\eta^i(z)$ and $\ov\eta^i(\ov z)$. 
This provides  a strong argument  
that the right-hand side of \eqref{hannover96} 
can be identified with the  $H$-field, which is also odd under the world-sheet parity.

Now, in case that the $H$-flux is the source of the NCA structure, the generalization
of our result \eqref{result} to any coordinate system is then given by
\begin{equation}
\label{threealggg}
   \bigl[X^i,X^j,X^k\bigr]= - \frac{\pi^2}{  \sqrt 8}\,   (\alpha')^2\,
 \bigl({\cal T^+}\bigr)^{ijk}=+\frac{\pi^2}{ \sqrt 8}\,   (\alpha')^2 
\, H^{ijk}\; ,
\end{equation}   
where the torsion $\mathcal T^+$ was defined in \eqref{torsion_458} and where we have employed that $f^{abc} = -\sqrt{k/2}\, H^{abc}$.
Furthermore, in this case the generalized 
Moyal-Weyl star-product reads
\eq{
\label{threebracket}
   F\tri G\tri H = \exp\left[ {\textstyle \frac{\pi^2}{12\sqrt{2}}} (\alpha')^2  H^{ijk}\,
      \partial^x_{i}\,\partial^y_{j}\,\partial^z_{k} \right]\, F(x)\, G(y)\,
   H(z)\Bigr|_{x=y=z} \;,
}
and one can again envision that 
 higher order $\alpha'$-corrections  of the $H$-flux are
captured by the NCA three-product \eqref{threebracket}.

\bigskip
Let us now discuss in some more detail the implications
of a non-trivial three-bracket \eqref{threealggg}.
We first recall that the computation leading to  \eqref{result} was done in the framework of conformal field theory and that we did not  refer to a specific embedding
of the WZW model into string theory.  
In particular,  this conformal field theory
can be part of a bosonic as well as of a supersymmetric string theory compactification, and
we expect in both cases 
that a non-vanishing $H$-flux leads to a non-trivial three-bracket.
Furthermore, type IIA superstring theory is
related to eleven-dimensional M-theory compactified on a
circle. When uplifting the three-bracket  structure
to M-theory, we can  expect a non-vanishing $G_4$-form flux
to induce a four-bracket of the form
\begin{equation}
[ X^i,X^j,X^k,X^l]\simeq {G}^{ijkl}\; .
\end{equation}
Following our argument from the introduction, this is consistent 
with the fact that on a probe membrane
one has to specify four-points to define an orientation.
Again, it is tempting to introduce a four-product 
\eq{
F\quat G\quat H \quat K
}
defined in complete  analogy
to the three-product  \eqref{threebracket}. 
Now, going back to the type IIA superstring by compactifying M-theory on a circle, 
the $G_4$-flux splits into $H$-flux as well as into 
R-R $F_4$-form flux. Therefore, we not only expect a non-vanishing three-bracket for the $H$-flux
but in addition also a non-vanishing four-bracket corresponding to $F_4$-flux.
Similarly, since M-theory compactified
on a  two-torus of vanishing volume
is dual to the type IIB superstring, we expect  a three-bracket
for both the $H$-flux and the R-R $F_3$-form flux.
However, note that the effect of these R-R fluxes cannot be detected by
 conformal field theory computations.

To summarize, if the origin of the three-bracket \eqref{result} is the $H$-flux, string
dualities  imply that
R-R fluxes in superstring theories
and  four-form flux in M-theory also induce
non-trivial three- and four-brackets for target space
coordinates.  


\subsubsection*{Generalized  Three-bracket Geometry}

Another intriguing possibility is to follow the idea
of {\it generalized} geometry and  keep the left- and right-moving coordinates
completely separated, leading to two NCA geometries. 
Indeed, for the geometric coordinates 
our results are also compatible with the following two three-bracket 
relations
\begin{equation}
\label{threebrackgeotwo}
\arraycolsep2pt
\begin{array}{lll}
\displaystyle  \bigl[X^i,X^j,X^k\bigr]_{\mbox{\scriptsize hol}} &
\displaystyle =
   \frac{\pi^2}{ 2\sqrt 2}\,   (\alpha')^2  \, \left( \omega^{ijk}+{\textstyle{\frac{1}{2}}}H^{ijk} \right)
& \displaystyle   =\frac{\pi^2}{ 2\sqrt 2}\,   (\alpha')^2  \, \Omega^{+\, ijk}\; , \\[4mm]
\displaystyle \bigl[\ov X^i,\ov X^j,\ov X^k\bigr]_{\mbox{\scriptsize anti-hol}} 
&\displaystyle =
   \frac{\pi^2}{ 2\sqrt 2}\,   (\alpha')^2  \, \left( \omega^{ijk}-{\textstyle {\frac{1}{ 2}}}H^{ijk} \right)
&\displaystyle   =\frac{\pi^2}{ 2\sqrt 2}\,   (\alpha')^2  \, \Omega^{-\, ijk}\; . 
\end{array}
\end{equation}
Instead of adding or subtracting  these two equations from the very beginning, as it was discussed below \eqref{threebrackgeo} or done in  \eqref{hannover96}, 
the new proposal is to first define two separate  NCA gravity theories and combine them only at the very end.
Thus, instead of a single three-product one is led to two products of the form
\eq{
\label{threebrackettwo}
   F\tri^{\!\!+} G\tri^{\!\! +} H &= \exp\left[ {\textstyle \frac{\pi^2}{12\sqrt{2}}} (\alpha')^2 \, \Omega^{+\, ijk}\,
      \partial^x_{i}\,\partial^y_{j}\,\partial^z_{k} \right]\, F(x)\, G(y)\,
   H(z)\Bigr|_{x=y=z} \;, \\
    F\tri^{\!\! -} G\tri^{\!\! -} H &= \exp\left[ {\textstyle \frac{\pi^2}{12\sqrt{2}}} (\alpha')^2\,  \Omega^{-\, ijk}\,
      \partial^x_{i}\,\partial^y_{j}\,\partial^z_{k} \right]\, F(x)\, G(y)\,
   H(z)\Bigr|_{x=y=z} \;,
}
out of which one might construct  two NCA Einstein-Hilbert like
actions $S_{\rm EH}(\;\tri^{\!\!+})$ and  $S_{\rm EH}(\;\tri^{\!\!-})$
(plus NCA deformations of the tree-level actions for the $B_2$-form
and the dilaton)
in analogy to noncommutative gauge theory.
The full string theoretical gravity theory should then be given by 
adding these two actions
\eq{
\label{einbert}
  S= S_{\rm EH}(\;\tri^{\!\!+})+S_{\rm EH}(\;\tri^{\!\!-})
   \sim\int \hat R(\;\tri^{\!\!+}) + \hat R(\;\tri^{\!\!-}) + \ldots \; .
}
The crucial question  is whether expanding such a putative 
{\it generalized  nonassociative gravity theory}  in terms of  the fundamental fields $(g,H,\Phi)$ can be equivalent to the $\alpha'$-expansion of the effective string theory action.  

Finally, we note that when uplifting this structure to M-theory, the natural expectation would be 
to obtain
both a three-product  $(\cdot \,\tri\, \cdot\, \tri\, \cdot)_{\omega}$
determined
by the geometric connection $\omega^{ijk}$ as well as a four-product
$(\cdot\, \quat\, \cdot\, \quat \,\cdot \, \quat \,\cdot)_{G}$ determined
by the four-form flux $G^{ijkl}$.


\subsubsection*{Four-bracket}

As emphasized above, one of our arguments for a non-trivial three-bracket structure is that the 
equal-time, equal-position cyclic double commutator of three fields $X^a(\sigma,\tau)$ 
is independent of the world-sheet 
coordinates $\sigma$ and $\tau$.
Motivated by this reasoning, let us  try to define other objects
which are independent of the world-sheet coordinates. 
For instance,  since the curvature $R_{ijkl}$ has
four indices, it is conceivable that it appears in  a four-bracket of the form
\eq{
\label{fourbracket}
   \bigl[X^i,X^j\,;\,X^k,X^l\bigr]\simeq (\alpha')^3\,  {R}^{ijkl} \;.
}
Note that this four-bracket is not completely anti-symmetric, but 
a priori only has  to reflect the symmetries of the curvature tensor.
One way to define such a four-bracket as an equal-time, equal-position three-commutator is the following
\begin{equation}
\label{quarticid}
\begin{split}
\bigl[X^a,X^b\,;\,X^c,X^d\bigr]= \lim_{\sigma_i\to \sigma}\quad 
&\Bigl[\bigl[[X^a(\sigma_1,\tau),\underbracket{\;X^b(\sigma_2,\tau)],X^c(\sigma_3,\tau)\bigr],X^d(\sigma_4,\tau)\Bigr]\;}\\[-0.5mm]
-&\Bigl[\bigl[[X^a(\sigma_1,\tau),\underbracket{\;X^b(\sigma_2,\tau)],X^d(\sigma_4,\tau)\bigr],X^c(\sigma_3,\tau)\Bigr]\;}\\[1.5mm]
-&\bigl[X^b X^a|X^c X^d\bigr] + \bigl[X^b X^a|X^d X^c\bigr] + \bigl[X^c X^d|X^a X^b\bigr] \\[0.5mm]
-& \bigl[X^d X^c|X^a X^b\bigr]-\bigl[X^c X^d|X^b X^a\bigr] + \bigl[X^d X^c|X^b
X^a\bigr] ,
\end{split}\hspace{-11pt}
\end{equation}
where the underbracket stands for taking a cyclic sum, and we have employed a short-hand notation for the last six  triple commutators.
We have evaluated \eqref{quarticid} for our example from section
\ref{sec:closed} and, neglecting the zero modes $x^a_0$, found 
\begin{equation}
\begin{split}
\bigl[X^a,X^b\,;\,X^c,X^d\bigr]&= \lim_{\sigma_i\to \sigma}  \;
 \frac{2}{k}\, \Bigl( f^{ab}{}_{u}\,  f^{ucd} + f^{ca}{}_{u}\,  f^{ubd} +
  f^{bc}{}_{u}\,  f^{uad}\Bigr) \: F(\sigma_i,\tau)\\[2mm]
&= \lim_{\sigma_i\to \sigma} \; \bigl({\cal R^+}\bigr)^{abcd} \; F(\sigma_i,\tau)=0 \; ,
\end{split}
\end{equation}
where  $F(\sigma_i,\tau)$ is an expression depending on the world-sheet
coordinates $\sigma_i$ and $\tau$.
In the second line we have employed \eqref{curvbb} and that 
the curvature $\mathcal R^+$ vanishes due to the Jacobi-identity. 
Therefore, from this conformal field theory  analysis we cannot decide whether the four-bracket is
generically zero or only vanishes on-shell.

\medskip
We note that it would be interesting to find more objects of this kind,
involving for instance  not just the
target space coordinates $X^a$ but also the target space momentum $P^a$.


\subsubsection*{Uncertainty Relation}

To conclude this section, let us comment on the physical implications
of  the non-vanishing three-bracket \eqref{result}. 
Similar to  a non-vanishing commutator
in quantum mechanics, we expect  a three-bracket
to imply a generalized uncertainty relation.  
In the literature (for instance in \cite{SheikhJabbari:2007iy,Ho:2007vk}) it was suggested that a three-bracket
$[X,Y,Z]= \ell_s^3$ implies a   triple    uncertainty relation of the form
\begin{equation}
\label{ur_lit}
\Delta X\,\Delta Y\,\Delta Z\, \ge \ell_s^3 \; .
\end{equation}
In appendix \ref{app_ucp},  we  made an attempt to generalize the derivation  
of the quantum mechanical
uncertainty relation $\Delta p\, \Delta x>\hbar/2$ to the present case.
We have obtained a result different from \eqref{ur_lit}, in particular we found
\begin{equation}
\begin{split}
&\Delta(X|Y)^2\, \Delta Z^2 + \Delta(Y|Z)^2\, \Delta X^2 +
\Delta(Z|X)^2\, \Delta Y^2 \ge \frac{\ell_s^6}{4} \;,
\end{split}
\end{equation}
where $\Delta(X|Y)^2$ denotes the uncertainty  of the commutator $i[X,Y]$ and similarly for the others
(see appendix \ref{app_ucp} for more details).
This means  that a non-vanishing three-bracket leads to 
an intertwined uncertainty relation for the position operators and
their mutual commutators.
Let us observe that this is consistent with our remark 4 
on page \pageref{remark_4}, which
stated that the single commutator of two target space coordinates is not a well-defined
fundamental object.  Furthermore, it would be interesting to generalize this
computation to the case of a fundamental non-vanishing four-bracket.


\section{Conclusions}
\label{sec:concl}

The central concern  of this paper was the study of
$n$-bracket structures for  target space coordinates as probed
by a closed string. For this purpose, we have studied in detail closed strings moving on the three-sphere $S^3$ in the presence of background $H$-flux. 
In particular, employing the exact solvability  of the $SU(2)_k$ WZW model
we have performed a conformal field theory computation of the equal-time,
equal-position cyclic double commutator of the fields $X^a$. 
Remarkably, despite the fact that the generators of the 
Kac-Moody algebra appearing in the mode expansion of $X^a$ 
satisfy a Jacobi-identity, we obtained  
a non-vanishing expression  independent of the
world-sheet coordinates. Therefore, we interpreted this result as an
indication for  a non-trivial three-bracket of the
target space coordinates.
However, 
in the course of the computation we made one technical assumption,
and the identification
of the  source for the non-trivial three-bracket
deserve further investigation and clarification.

Motivated by our findings, analogous to the appearance of  a deformed
bi-product in the open string sector (i.e. for a gauge theory), we have 
proposed to
introduce a deformed three-product for the target space coordinates 
in the closed string sector (i.e. for the gravity theory).
For  a non-vanishing fundamental three-bracket, we also made
a new proposal for the implied uncertainty relation.
This  was supported by an explicit generalization of
the quantum mechanical derivation of the original Heisenberg
uncertainty principle.

If a non-trivial three-bracket/three-product 
structure is indeed present in the closed sector of string theory,
new  conceptional questions arise both from a physical and mathematical  point of view:
\begin{itemize}

\item Is it possible to reconstruct the string equations of motion, including
all $\alpha'$-corrections,  in a pure target space approach via  an 
Einstein-like gravity theory on a nonassociative space-time?

\item If the target space-time is NCA, how does the non-linear sigma-model 
of a string moving in such a background  take this structure into account?

\item To our knowledge the mathematical foundations of  NCA geometries
are far less developed than for noncommutative but associative \pagebreak[2]
geometries. If indeed   quantum gravity requires such a 
 framework, can a mathematically
rigorous NCA geometry incorporating  a deformation or generalization of
 general covariance  be developed.

\end{itemize}
Clearly, more work is needed to support or disprove
the assumption and proposals  made in  this
paper, and to  advance in the directions  proposed.


\subsubsection*{Acknowledgements}
We would like to thank 
Andreas Deser and Dieter L\"ust for 
valuable comments and discussions.
We furthermore thank Arthur Hebecker, Igor Khavkine and Timo Weigand for discussions on an earlier version of this article.
This work was initiated  during the workshop {\it Strings
at the LHC and in the Early Universe} held in spring 2010 at the
Kavli Institute for Theoretical Physics. We would like to
thank the institute for their hospitality and for creating
such a pleasant research environment. E.P. also thanks
the Max-Planck-Institut f\"ur Physik in Munich for hospitality. 
This research was supported in part by the National Science Foundation under Grant No. PHY05-51164.


\newpage

\appendix

\section{Generalized Uncertainty Principle}
\label{app_ucp}

An important general question is what kind of  uncertainty 
relation originates from a non-vanishing  three-bracket of the form
\begin{equation}
  \label{resu_app}
      [X,Y,Z]= \ell_s^3 \;.
\end{equation}
Here we present a possible  self-consistent derivation,
which follows closely the quantum mechanical derivation
of the Heisenberg uncertainty relation.

In  \cite{Minic:2002pd} it was noted  that in a theory with a
quantum Nambu three-bracket it may be useful to work on a Hilbert-space 
${\cal H}_{\mathbb H}$ defined
over the quaternions $\mathbb{H}$. Here, we also expand a 
state $|\psi\rangle\in {\cal H}_{\mathbb H}$ as
\eq{
  |\psi\rangle = |\psi_0\rangle + i |\psi_1\rangle  + j|\psi_2\rangle  + k|\psi_3\rangle \;,
}
with $|\psi_i\rangle \in {\cal H}_{\mathbb R}$. Recall that the symbols $i$, $j$ and $k$ satisfy the usual quaternionic relations $i^2=j^2=k^2=-1$, $ij=k$, $jk=i$, and so on. 
The conjugate state is defined as
$\langle\psi| = \langle\psi_0| -i \langle\psi_1|  - j\langle\psi_2| - k\langle \psi_3|$, 
and the scalar product therefore reads 
\begin{equation}
\begin{split}
 \langle \phi|\psi\rangle=  \sum_{a=0}^3  \langle \phi_a| \psi_a\rangle
    + i\,& \Bigl( \langle \phi_0 |\psi_1\rangle - \langle \phi_1
 |\psi_0\rangle +\langle \phi_3 |\psi_2\rangle- \langle \phi_2 |\psi_3\rangle  \Bigr)\\[-2.5mm]
     +j\,& \Bigl( \langle \phi_0 |\psi_2\rangle - \langle \phi_2
 |\psi_0\rangle + \langle \phi_1 |\psi_3\rangle - \langle \phi_3 |\psi_1\rangle \Bigr)\\
     +k\, &\Bigl( \langle \phi_0 |\psi_3\rangle - \langle \phi_3
 |\psi_0\rangle + \langle \phi_2 |\psi_1\rangle  - \langle \phi_1 |\psi_2\rangle \Bigr)\; .
\end{split}
\end{equation}
The starting point of our consideration is a state $A|\psi\rangle$, where the operator $A$ is chosen to have the following form 
\begin{equation}
\label{state}
 A= i\hspace{0.5pt}\lambda_1 X + j\hspace{0.5pt}\lambda_2 Y + k\hspace{0.5pt} \lambda_3 Z +
      k\hspace{0.5pt}  \lambda_1 \lambda_2 [X,Y] 
      +  j\hspace{0.5pt}  \lambda_1 \lambda_3 [Z,X]+
     i\hspace{0.5pt}  \lambda_2 \lambda_3 [Y,Z] \;,
\end{equation}
with $\lambda_a\in \IR$. 
Note that in \eqref{state}, six self-adjoint operators appear which can be interpreted as observables
\begin{equation}
\label{selfad}
\Bigl\{ X\,,\,Y\,,\,Z\,,\,i[Y,Z]\,,\,j[Z,X]\,,\,k[X,Y]\Bigr\} \;.
\end{equation}
Now, we are  going to employ the  three-bracket $[X,Y,Z]=[X,[Y,Z]]+{\rm cycl.} =\ell_s^3$ in the evaluation of the condition
\eq{
  \label{pos_con}
  \langle A\psi \,|\, A\psi\rangle\ge 0 \;,
}  
and analyze the resulting expression.
However,
to simplify our discussion, we assume the expectation values for the operators \eqref{selfad} to vanish, that is
\begin{equation}
\label{vanishvev}
\begin{split}
  &0=\bigl\langle \psi\bigr| X \bigl|\psi\bigr\rangle
  = \bigl\langle \psi\bigr|Y \bigl|\psi\bigr\rangle
  = \bigl\langle \psi\bigr|Z \bigl|\psi\bigr\rangle \;, \\[2mm]
  &0=\bigl\langle \psi\bigr| k[X,Y] \bigl|\psi\bigr\rangle
  = \bigl\langle \psi\bigr| j [Z,X] \bigl|\psi\bigr\rangle
  =\bigl\langle \psi\bigr| i [Y,Z] \bigl|\psi\bigr\rangle  \; .
\end{split}
\end{equation}
Note that the second line is consistent with our point of view that
the three-bracket is the fundamental structure of the problem, 
and thus no non-trivial uncertainty relation for only two
position operators should appear.
Furthermore, it turns out that we have to require in addition 
that some operators  appearing
in \eqref{state} can be measured simultaneously, which leads to
\begin{equation}
\label{vanishvevb}
\begin{split}
  0= \bigl\langle \psi\bigr| [[X,Y],[Z,X]] \bigl|\psi\bigr\rangle 
  =\bigl\langle \psi\bigr|
  [[X,Y],[Y,Z]] \bigl|\psi\bigr\rangle 
  =\bigl\langle \psi\bigr| [[Z,X],[Y,Z]] \bigl|\psi\bigr\rangle  .
\end{split}
\end{equation}
When evaluating the expression \eqref{pos_con}, different  terms appear.
In particular, from the product of two linear terms in \eqref{state} we find
\begin{equation}
\label{lineara}
\begin{split}
 \langle A\psi|A\psi\rangle_{11}&=\lambda_1^2\, \langle \psi| X^2 |\psi\rangle+
    \lambda_2^2\, \langle \psi| Y^2 |\psi\rangle+
    \lambda_3^2\,  \langle \psi| Z^2 |\psi\rangle \\ 
   &=\lambda_1^2\, \Delta X^2+\lambda_2^2\, \Delta Y^2+\lambda_3^2\,
    \Delta Z^2\, .
\end{split}
\end{equation}
The product of one linear and one quadratic term in \eqref{state} leads to
\eq{
\label{linquada}
 \langle A\psi|A\psi\rangle_{12}&=
  \lambda_1\lambda_2\lambda_3 \,\langle \psi|\, [X,[Y,Z]]+[Y,[Z,X]]+[Z,[X,Y]]\,|\psi\rangle\\
   &\quad-k \lambda^2_1\lambda_3\, \langle \psi|\, \{X,[Z,X]\}\,|\psi\rangle +
    j \lambda^2_1\lambda_2\, \langle \psi|\, \{X,[X,Y]\}\,|\psi\rangle 
   +{\rm cycl.}\\
&=  \lambda_1\lambda_2\lambda_3\,  \ell_s^3 \;, \\[-3.5mm]
}
where in the last line we have employed \eqref{resu_app} and also
required the vanishing of the expectation values 
$\langle \psi|\, \{X,[Z,X]\}\,|\psi\rangle$, $\langle \psi|\,
\{X,[X,Y]\}\,|\psi\rangle$ as well as their cyclic permutations.
The interpretation of these conditions is not clear to us.
Finally, multiplying two quadratic terms in  \eqref{state}, we arrive at
\begin{equation}
\label{quada}
\begin{split}
 \langle A\psi|A\psi\rangle_{22}&=\lambda_1^2\lambda_2^2
\, \langle \psi| (k[X,Y])^2 |\psi\rangle+
\lambda_1^2\lambda_3^2
\, \langle \psi| (j[Z,X])^2 |\psi\rangle\\
&\hspace{124.5pt}+
\lambda_2^2\lambda_3^2
\, \langle \psi| (i[Y,Z])^2 |\psi\rangle\\[0.1cm]
&=\lambda_1^2\lambda_2^2
\, \Delta(X|Y)^2 + \lambda_1^2\lambda_3^2
\, \Delta(Z|X)^2+
\lambda_2^2\lambda_3^2
\,\Delta(Y|Z)^2 \;,
\end{split}
\end{equation}
where we utilized \eqref{vanishvevb} and introduced the notation
 $\Delta(X|Y)^2=\langle \psi| (k[X,Y])^2 |\psi\rangle$, and similarly for the others. 
Adding all three contributions, we arrive at 
\begin{equation}
\begin{split}
 0\le \lambda_1^2\lambda_2^2
\, \Delta(X|Y)^2 +& \lambda_1^2\lambda_3^2
\, \Delta(Z|X)^2+
\lambda_2^2\lambda_3^2
\,\Delta(Y|Z)^2 \\[1.5mm]
 +\, \lambda_1\lambda_2\lambda_3\,  \ell_s^3+&
\lambda_1^2\, \Delta X^2+\lambda_2^2\, \Delta Y^2+\lambda_3^2\,
    \Delta Z^2 \;.
\end{split}
\end{equation}
This relation has to be true for all values of  $\lambda_1,\lambda_2,\lambda_3\in \mathbb R$, which is satisfied if and only if the following
uncertainty relation holds
\begin{equation}
\label{ur_final}
\begin{split}
&\Delta(X|Y)^2\, \Delta Z^2 + \Delta(Y|Z)^2\, \Delta X^2 +
\Delta(Z|X)^2\, \Delta Y^2 \ge \frac{\ell_s^6}{  4}\; .
\end{split}
\end{equation}
Note that due to the assumptions made in the course of this derivation,
equation \eqref{ur_final} is to be supplemented by the trivial uncertainty relations
\begin{equation}
\arraycolsep2pt
\begin{array}{@{}l@{\hspace{34pt}}l@{\hspace{34pt}}l@{}}
0\leq\Delta X\, \Delta Y \;, & 
0\leq\Delta X\, \Delta Z \;, &
0\leq\Delta Y\, \Delta Z \;, \\[1mm]
0\leq\Delta(X|Y)\, \Delta(Y|Z)\;, &
0\leq\Delta(Y|Z)\, \Delta(Z|X)\;, &
0\leq\Delta(Z|X)\, \Delta(X|Y)\; .
\end{array}
\end{equation}
Therefore, we have arrived at the result that a fundamental non-vanishing
three-bracket gives rise to an uncertainty relation involving not
only the positions operators but also their commutators.


\clearpage
\nocite{*}
\bibliography{rev}  

\providecommand{\href}[2]{#2}\begingroup\raggedright\begin{thebibliography}{10}

\bibitem{Connes:1997cr}
A.~Connes, M.~R. Douglas, and A.~S. Schwarz, ``{Noncommutative geometry and
  matrix theory: Compactification on tori},'' {\em JHEP} {\bf 02} (1998) 003,
\href{http://www.arXiv.org/abs/hep-th/9711162}{{\tt hep-th/9711162}}.

\bibitem{Chu:1998qz}
C.-S. Chu and P.-M. Ho, ``{Noncommutative open string and D-brane},'' {\em
  Nucl. Phys.} {\bf B550} (1999) 151--168,
\href{http://www.arXiv.org/abs/hep-th/9812219}{{\tt hep-th/9812219}}.

\bibitem{Schomerus:1999ug}
V.~Schomerus, ``{D-branes and deformation quantization},'' {\em JHEP} {\bf 06}
  (1999) 030,
\href{http://www.arXiv.org/abs/hep-th/9903205}{{\tt hep-th/9903205}}.

\bibitem{Seiberg:1999vs}
N.~Seiberg and E.~Witten, ``{String theory and noncommutative geometry},'' {\em
  JHEP} {\bf 09} (1999) 032,
\href{http://www.arXiv.org/abs/hep-th/9908142}{{\tt hep-th/9908142}}.

\bibitem{Frohlich:1993es}
J.~Fr{\"o}hlich and K.~Gawedzki, ``{Conformal field theory and geometry of
  strings},''
\href{http://www.arXiv.org/abs/hep-th/9310187}{{\tt hep-th/9310187}}.

\bibitem{Lust:2010iy}
D.~L{\"u}st, ``{T-duality and closed string non-commutative (doubled)
  geometry},''
\href{http://www.arXiv.org/abs/1010.1361}{{\tt 1010.1361}}.

\bibitem{Hoppe:1996xp}
J.~Hoppe, ``{On M-Algebras, the Quantisation of Nambu-Mechanics, and Volume
  Preserving Diffeomorphisms},'' {\em Helv. Phys. Acta} {\bf 70} (1997)
  302--317,
\href{http://www.arXiv.org/abs/hep-th/9602020}{{\tt hep-th/9602020}}.

\bibitem{Ho:2007vk}
P.-M. Ho and Y.~Matsuo, ``{A toy model of open membrane field theory in
  constant 3- form flux},'' {\em Gen. Rel. Grav.} {\bf 39} (2007) 913--944,
\href{http://www.arXiv.org/abs/hep-th/0701130}{{\tt hep-th/0701130}}.

\bibitem{Chu:2009iv}
C.-S. Chu and D.~J. Smith, ``{Towards the Quantum Geometry of the M5-brane in a
  Constant $C$-Field from Multiple Membranes},'' {\em JHEP} {\bf 04} (2009)
  097,
\href{http://www.arXiv.org/abs/0901.1847}{{\tt 0901.1847}}.

\bibitem{Bagger:2006sk}
J.~Bagger and N.~Lambert, ``{Modeling multiple M2's},'' {\em Phys. Rev.} {\bf
  D75} (2007) 045020,
\href{http://www.arXiv.org/abs/hep-th/0611108}{{\tt hep-th/0611108}}.

\bibitem{Witten:1983ar}
E.~Witten, ``{Nonabelian bosonization in two dimensions},'' {\em Commun. Math.
  Phys.} {\bf 92} (1984)
455--472.

\bibitem{Gepner:1986wi}
D.~Gepner and E.~Witten, ``{String Theory on Group Manifolds},'' {\em Nucl.
  Phys.} {\bf B278} (1986)
493.

\bibitem{Callan:1986bc}
C.~G. Callan, Jr., C.~Lovelace, C.~R. Nappi, and S.~A. Yost, ``{String Loop
  Corrections to beta Functions},'' {\em Nucl. Phys.} {\bf B288} (1987)
525.

\bibitem{Abouelsaood:1986gd}
A.~Abouelsaood, C.~G. Callan, Jr., C.~R. Nappi, and S.~A. Yost, ``{Open Strings
  in Background Gauge Fields},'' {\em Nucl. Phys.} {\bf B280} (1987)
599.

\bibitem{Alekseev:1999bs}
A.~Y. Alekseev, A.~Recknagel, and V.~Schomerus, ``{Non-commutative world-volume
  geometries: Branes on SU(2) and fuzzy spheres},'' {\em JHEP} {\bf 09} (1999)
  023,
\href{http://www.arXiv.org/abs/hep-th/9908040}{{\tt hep-th/9908040}}.

\bibitem{Cornalba:2001sm}
L.~Cornalba and R.~Schiappa, ``{Nonassociative star product deformations for
  D-brane worldvolumes in curved backgrounds},'' {\em Commun. Math. Phys.} {\bf
  225} (2002) 33--66,
\href{http://www.arXiv.org/abs/hep-th/0101219}{{\tt hep-th/0101219}}.

\bibitem{Herbst:2001ai}
M.~Herbst, A.~Kling, and M.~Kreuzer, ``{Star products from open strings in
  curved backgrounds},'' {\em JHEP} {\bf 09} (2001) 014,
\href{http://www.arXiv.org/abs/hep-th/0106159}{{\tt hep-th/0106159}}.

\bibitem{Callan:1991at}
C.~G. Callan, Jr., J.~A. Harvey, and A.~Strominger, ``{Supersymmetric string
  solitons},''
\href{http://www.arXiv.org/abs/hep-th/9112030}{{\tt hep-th/9112030}}.

\bibitem{Khuri:1990mg}
R.~R. Khuri, ``{Some instanton solutions in string theory},'' {\em Phys. Lett.}
  {\bf B259} (1991)
261--266.

\bibitem{Braaten:1985is}
E.~Braaten, T.~L. Curtright, and C.~K. Zachos, ``{Torsion and Geometrostasis in
  Nonlinear Sigma Models},'' {\em Nucl. Phys.} {\bf B260} (1985)
630.

\bibitem{Nappi:1993ie}
C.~R. Nappi and E.~Witten, ``{A WZW model based on a nonsemisimple group},''
  {\em Phys. Rev. Lett.} {\bf 71} (1993) 3751--3753,
\href{http://www.arXiv.org/abs/hep-th/9310112}{{\tt hep-th/9310112}}.

\bibitem{Dolan:2002px}
L.~Dolan and C.~R. Nappi, ``{Noncommutativity in a time-dependent
  background},'' {\em Phys. Lett.} {\bf B551} (2003) 369--377,
\href{http://www.arXiv.org/abs/hep-th/0210030}{{\tt hep-th/0210030}}.

\bibitem{deser}
R.~Blumenhagen, A.~Deser, D.~L{\"u}st, and E.~Plauschinn, ``work in
  progress,''.

\bibitem{Rey:1991uu}
S.-J. Rey, ``{On string theory and axionic strings and instantons},''.
  Presented at Particle and Fields '91 Conf., Vancouver, Canada, Aug 18-22,
  1991.

\bibitem{Aschieri:2005yw}
P.~Aschieri {\em et al.}, ``{A gravity theory on noncommutative spaces},'' {\em
  Class. Quant. Grav.} {\bf 22} (2005) 3511--3532,
\href{http://www.arXiv.org/abs/hep-th/0504183}{{\tt hep-th/0504183}}.

\bibitem{SheikhJabbari:2007iy}
M.~M. Sheikh-Jabbari, ``{An N-tropic solution to the cosmological constant
  problem},''
\href{http://www.arXiv.org/abs/hep-ph/0701084}{{\tt hep-ph/0701084}}.

\bibitem{Minic:2002pd}
D.~Minic and H.~C. Tze, ``{Nambu Quantum Mechanics: A Nonlinear Generalization
  of Geometric Quantum Mechanics},'' {\em Phys. Lett.} {\bf B536} (2002)
  305--314,
\href{http://www.arXiv.org/abs/hep-th/0202173}{{\tt hep-th/0202173}}.

\bibitem{Nambu:1973qe}
Y.~Nambu, ``{Generalized Hamiltonian dynamics},'' {\em Phys. Rev.} {\bf D7}
  (1973)
2405--2414.

\end{thebibliography}\endgroup
\bibliographystyle{utphys}


\end{document}